# Symmetries of Ginsparg-Wilson Chiral Fermions

Jeffrey E. Mandula[*]

University of Washington, Department of Physics, Seattle, WA 98105, USA

**Abstract**

The group structure of the variant chiral symmetry discovered by Lüscher in the Ginsparg-Wilson description of lattice chiral fermions is analyzed. It is shown that the group contains an infinite number of linearly independent symmetry generators, and the Lie algebra is given explicitly. CP is an automorphism of this extended chiral group, and the CP transformation properties of the symmetry generators are found. The group has an infinite-parameter invariant subgroup, and the factor group, whose elements are its cosets, is isomorphic to the continuum chiral symmetry group. Features of the currents associated with these symmetries are discussed, including the fact that some different, non-commuting symmetry generators lead to the same Noether current. These are universal features of lattice chiral fermions based on the Ginsparg-Wilson relation; they occur in the overlap, domain-wall, and perfect-action formulations. In a solvable example, free overlap fermions, these non-canonical elements of lattice chiral symmetry are related to complex energy singularities that violate reflection positivity and impede continuation to Minkowski space.

---

[*] E-mail address:  mandula@post.harvard.edu

I. **INTRODUCTION**

This article is concerned with the structure of the symmetry group of lattice chiral fermions based on the Ginsparg-Wilson relation [1]. Several independent constructions of lattice chiral fermions that incorporated the Ginsparg-Wilson relation were developed in the 1990's [2,3,4], although the fact that the Ginsparg-Wilson relation was the key to how they all represented chiral symmetry was realized only afterwards. The transformations which, acting on the fermion variables, left the fermion action invariant were subsequently discovered by Lüscher [5]. It is curious that this timeline of developments is almost the opposite to how the consequences of symmetries are usually found. There have been several excellent reviews of these developments in the Proceedings of International Symposia on Lattice Field Theory from that time [6].

In a recent note [7] it was pointed out that lattice chiral symmetry based on the Ginsparg-Wilson relation has several surprising features. They follow from the fact that there are multiple asymmetric forms of Lüscher's symmetry transformation. The different forms are mathematically inequivalent, and, in fact, the symmetry generators corresponding to them do not even commute with one another. As a consequence, it was observed that the group of lattice chiral symmetries is not a finite-dimensional Lie group, but has an infinite number of linearly independent generators. CP is an automorphism of this extended group, but in general it mixes generators. A quite peculiar feature of Ginsparg-Wilson chiral symmetry is that the usual, one-to-one correspondence between currents and symmetries is lost. Some of the different, non-commuting symmetries lead to identical Noether currents.

Underlying these strange features is a key difference between the Euclidean space path integral formulation of field theory and canonical field theory: that in the path integral fermion and anti-fermion variables must be treated as independent, and not conjugate variables [8]. Indeed, Lüscher's symmetry transformation quite explicitly exploits this characteristic of Euclidean fermionic path integrals.

In this article, we examine the structure of lattice chiral symmetry more fully, and we clarify the roles of the "extra" symmetry transformations. We display the structure of the extended chiral group and show that the "extra" transformations correspond to an invariant subgroup of the extended group. From a group theoretic perspective, the fact that this subgroup is not trivial is implicated in all of the above peculiarities of Ginsparg-Wilson fermions. The beginning sections of this article follow the order of ref. [7], extending the treatment to include flavored as well as flavor-singlet transformations.

In Sec. II, we show that the different forms of Lüscher's asymmetric symmetry transformation do not commute with each other, and that this enlarges the chiral symmetry group of Ginsparg-Wilson fermions to an infinite-parameter Lie group. The minimal extension needed in order that the algebra of symmetry generators close is constructed. We evaluate the action of CP on the generators of the group in Sec. III.

This much of the analysis depends only on the form of Lüscher's transformation, that is on the Ginsparg-Wilson relation itself.

In Sec. IV, following the procedure Kikukawa and Yamada [9], we construct the conserved Noether currents associated with each of the generators of the extended chiral group. In order to display the currents in closed form, an explicit construction of the chiral Ginsparg-Wilson action is needed, and we use overlap fermions based on the Wilson action. We show that the mismatch between the symmetry generators and the currents derived from them occurs for flavored generators as well as for flavor singlets. There are two multiplets of mathematically different transformations which give rise to identical Noether currents. We derive the Ward identities for the currents by the standard technique of making a change of integration variable corresponding to an infinitesimal space-time dependent version of each of the symmetry transformations in Sec. V.

In Sec. VI, we show that the extended lattice chiral group factors into the direct product of an infinite-parameter invariant subgroup and a finite-dimensional Lie group. We display the elements of the infinite-parameter subgroup and the commutators of their generators explicitly, and show that it is an invariant subgroup. We calculate the commutators of the generators of the factor group formed from the cosets of the infinite-parameter subgroup, and show that it is isomorphic to the ordinary continuum chiral group. CP is an automorphism of each of the factors of the full lattice chiral group separately. The generators of the infinite-parameter subgroup are not in general CP eigenstates. However, the generators of the factor group are CP eigenstates and transform under CP exactly as in the continuum. The elements of the factor group, being cosets and not individual group elements, are not directly represented on the fermions operators.

In Sec. VII, we address issues that arise in continuing Green's functions from Euclidean to Minkowski space. We examine the solvable example of free overlap fermions, which of course has all the properties discussed above. We focus on the behavior of the theory as we take the continuum limit. Taking the continuum limit only in time produces a theory with unphysical singularities in the complex energy plane. When the full continuum limit is taken, keeping the negative mass parameter proportional to the inverse lattice spacing, these singularities recede to infinity. That limit yields a valid continuum theory that has ordinary, conventional chiral symmetry. The infinite-parameter invariant subgroup that encapsulates the "extra" chiral symmetries becomes trivial, *i.e.* all of its generators vanish.

In Sec. VIII, we discuss the non-canonical aspects of Ginsparg-Wilson chiral fermions. The peculiar features of Ginsparg-Wilson fermions discussed above, specifically the multiple, asymmetric forms of Lüscher's chiral transformations, which causes the enlargement of the chiral symmetry group, and the mismatch between symmetries and their currents, could not occur in a canonically quantized field theory. They are the result of employing the freedom that comes with the need to treat fermions and anti-fermions as independent variables of integration in Euclidean fermionic path integrals. However, treatments of other field theories by path integrals do not lead to such peculiarities, even though in any Euclidean path integral, anti-fermion variables are independent of fermion variables. The difference between Ginsparg-Wilson fermions and other treatments of symmetry using Euclidean space path integrals is that one usually considers only symmetry transformations that act on the independent anti-fermion variables as they would if the anti-fermion variables were the conjugates of the fermion variables. However, none of the symmetries of Ginsparg-Wilson fermions, neither the forms discovered by Lüscher nor the extensions thereof discussed here, have this property.

In Sec. IX, we summarize the arguments and indicate that they form a coherent picture. Since all of these issues were uncovered by analyzing the symmetry group of the Ginsparg-Wilson equation, they are universal properties of Ginsparg-Wilson fermions and are inherent in all implementations of lattice fermions based on the Ginsparg-Wilson relation, including overlap, domain-wall, and perfect-action chiral fermions.

## II.  THE EXTENDED CHIRAL ALGEBRA

The following discussion uses only the Ginsparg-Wilson relation [1], the most familiar form of which (suppressing the lattice spacing) is

$$\gamma_5 D + D \gamma_5 = D \gamma_5 D \tag{1}$$

where $D$ is the lattice Dirac kernel, a $\gamma_5$-hermitian ($\gamma_5 D \gamma_5 = D^\dagger$) matrix labeled by color, spin, and lattice site indices. If we write the kernel as $D = 1 - V$, then $V$ is unitary (as well as $\gamma_5$-hermitian). In fact, these properties of $V$,

$$V^{-1} = V^\dagger = \gamma_5 V \gamma_5 \tag{2}$$

are completely equivalent to the Ginsparg-Wilson equation (1).

The (flavor singlet) chiral transformations under which the fermion action $S_F = \bar\psi D \psi$ is invariant are generated by [5]

$$\begin{aligned}\delta \psi &= \gamma_5 (1-D) \psi = \gamma_5 V \psi \\ \delta \bar\psi &= \bar\psi \gamma_5\end{aligned} \tag{3}$$

The asymmetric treatment of lattice fermions and anti-fermions is allowed because fermions and anti-fermions enter the Euclidean path integral as independent, not conjugate, variables. One may modify the transformation of the anti-fermion variables instead of the fermion variables,

$$\begin{aligned}\delta' \psi &= \gamma_5 \psi \\ \delta' \bar\psi &= \bar\psi(1-D)\gamma_5 = \bar\psi V \gamma_5\end{aligned} \tag{4}$$

or use any linear combination of these transformation rules. Whatever one's choice, the Ginsparg-Wilson equation insures the invariance of the action $S_F$.

These transformations are universally regarded as physically equivalent, but they are certainly different. Furthermore, each generates a symmetry of $S_F$. This means that the full chiral symmetry group of this theory is not generated by a choice of one or the other of these transformations (nor a linear combination thereof), but by both. This "enlarged" chiral group is not the end of the story either, because the two transformations do not even commute. Explicitly,

$$[\delta', \delta]\psi = \delta'\delta\psi - \delta\delta'\psi = (V^{-1} - V)\psi$$
$$[\delta', \delta]\bar\psi = \delta'\delta\bar\psi - \delta\delta'\bar\psi = \bar\psi(V - V^{-1}) \tag{5}$$

Their commutator, a vector rather than an axial transformation, is yet another symmetry of the fermion action. Upon further commutation, each of the chiral transformations (3) and (4) generates new symmetry transformations, further enlarging the chiral group.

A full group is easily found by writing the Ginsparg-Wilson equation as

$$\gamma_5 D + D \gamma_5 V = 0 \tag{6}$$

a form that displays the symmetry under Eq. (3) most clearly. Right multiplication by $V^{n-1}$ and the use of the unitarity and $\gamma_5$-hermiticity of $V$ gives

$$V^{1-n}\gamma_5 D + D\gamma_5 V^n = 0 \tag{7}$$

while right multiplication by $\gamma_5 V^{n+1}$, and noting that $\gamma_5 D \gamma_5 = -V^{-1}D$, gives

$$-V^n D + D V^n = 0 \tag{8}$$

From these we can read off the remaining axial and vector transformations of the fermion variables that generate symmetries of the Ginsparg-Wilson action. They are (including flavored transformations)

$$\begin{aligned}
\delta^{(n)}_{A,i}\psi &= \lambda_i \gamma_5 V^n \psi \\
\delta^{(n)}_{A,i}\bar{\psi} &= \bar{\psi} V^{1-n} \gamma_5 \lambda_i \\
\delta^{(n)}_{V,i}\psi &= -i\lambda_i V^n \psi \\
\delta^{(n)}_{V,i}\bar{\psi} &= i\bar{\psi} V^n \lambda_i
\end{aligned} \tag{9}$$

We recognize the two axial transformations with which we began this discussion as

$$\begin{aligned}
\delta &= \delta^{(1)}_{A,0} \\
\delta' &= \delta^{(0)}_{A,0}
\end{aligned} \tag{10}$$

The commutators of the generalized axial and vector generators are

$$\begin{aligned}
[\delta^{(n)}_{A,i},\delta^{(m)}_{A,j}] &= id_{ijk}\left(\delta^{(n-m)}_{V,k} - \delta^{(m-n)}_{V,k}\right) + f_{ijk}\left(\delta^{(n-m)}_{V,k} + \delta^{(m-n)}_{V,k}\right) \\
[\delta^{(n)}_{V,i},\delta^{(m)}_{A,j}] &= id_{ijk}\left(\delta^{(m-n)}_{A,k} - \delta^{(m+n)}_{A,k}\right) - f_{ijk}\left(\delta^{(m+n)}_{A,k} + \delta^{(m-n)}_{A,k}\right) \\
[\delta^{(n)}_{V,i},\delta^{(m)}_{V,j}] &= -2f_{ijk}\delta^{(n+m)}_{V,k}
\end{aligned} \tag{11}$$

This shows that the algebra closes and so further enlargement of the chiral group is not required. This minimal chiral group is of infinite rank, and the vector transformations $\delta^{(n)}_{V,i}$, with $i$ such that $\{\lambda_i\}$ are a complete set of commuting flavor matrices, are mutually commuting generators.

Infinite parameter symmetry groups are often a sign of some disease in a theory, and so a few remarks about this are in order. First of all, the elaborate structure of the lattice chiral symmetry group that we have described is built into the Ginsparg-Wilson equation, and applies to all implementations of the relation. The enlargement of the symmetry group, in fact the existence of the group itself, reflects $\psi$ and $\bar{\psi}$ being independent variables in the Euclidean space path integral.

## III. CP SYMMETRY

The only notable failure of the Ginsparg-Wilson implementation of chiral symmetry has been the inability, at least to date, to use it to construct lattice fermions with chiral interactions [10, 11]. (For recent alternative constructions of lattice chiral fermions see [12, 13, 14].) This failure is related to the fact that CP transforms the two forms of the asymmetric chiral transformations $\delta$ and $\delta'$ into one another [10]. However, we will see that CP is an automorphism of the enlarged chiral group.

Under parity, the fermion variables and $V = 1 - D$ transform as

$$\begin{aligned}
\mathcal{P}\, \psi &\to \gamma_4 P \psi \\
\mathcal{P}\, \bar{\psi} &\to \bar{\psi} P \gamma_4 \\
\mathcal{P}\, V &\to \gamma_4 P V P \gamma_4
\end{aligned} \tag{12}$$

where the matrix $P = P^{-1}$ acts only on the site labels and reflects the $x_{1,2,3}$ indices.

$$(P)_{(x_1,x_2,x_3,x_4),(y_1,y_2,y_3,y_4)} = \delta_{x_1,-y_1} \delta_{x_2,-y_2} \delta_{x_3,-y_3} \delta_{x_4,y_4} \tag{13}$$

Charge conjugation is represented on the link variables by complex conjugation, so the fermion variables and $V$ transform as

$$\begin{aligned}
\mathcal{C}\, \psi &\to C\, \bar{\psi}^\mathsf{T} \\
\mathcal{C}\, \bar{\psi} &\to -\psi^\mathsf{T} C^{-1} \\
\mathcal{C}\, V &\to \Gamma V^* \Gamma^{-1}
\end{aligned} \tag{14}$$

The matrices $C$ and $\Gamma$ depend on the representation of the Dirac matrices. $\Gamma$ flips the sign of the imaginary $\gamma$-matrices, so that

$$\Gamma \gamma_\mu^* \Gamma^{-1} = \gamma_\mu \tag{15}$$

and invariance of the action requires $C = \Gamma \gamma_5^\mathsf{T}$.

The CP transformation properties of the generators of the lattice chiral group follow from their action on the CP transformed variables. In a basis in which all the flavor matrices $\lambda_i$ are hermitian, they are

$$\begin{aligned}
(\mathcal{CP})^{-1} \delta_{A,i}^{(n)} (\mathcal{CP}) \, \psi &= -\lambda_i \gamma_5 V^{1-n} \psi \\
(\mathcal{CP})^{-1} \delta_{A,i}^{(n)} (\mathcal{CP}) \, \bar\psi &= -\bar\psi V^n \gamma_5 \lambda_i \\
(\mathcal{CP})^{-1} \delta_{V,i}^{(n)} (\mathcal{CP}) \, \psi &= i \lambda_i V^n \psi \\
(\mathcal{CP})^{-1} \delta_{V,i}^{(n)} (\mathcal{CP}) \, \bar\psi &= -i \bar\psi V^n \lambda_i
\end{aligned} \tag{16}$$

Comparing the above with the definitions of the generators (9) shows that

$$\begin{aligned}
(\mathcal{CP})^{-1} \delta_{A,i}^{(n)} (\mathcal{CP}) &= -\delta_{A,i}^{(1-n)} \\
(\mathcal{CP})^{-1} \delta_{V,i}^{(n)} (\mathcal{CP}) &= -\delta_{V,i}^{(n)}
\end{aligned} \tag{17}$$

Thus CP is seen to be an automorphism of the extended chiral symmetry group, but one that mixes the axial generators. This is the symmetry group view of the difficulty in reconciling CP invariance with chiral symmetry as realized through the Ginsparg-Wilson equation.

## IV.  EUCLIDEAN  SPACE  CURRENTS

The conserved currents associated with each of the $\delta_{A,i}^{(n)}$ and $\delta_{V,i}^{(n)}$ symmetries are constructed following the Noether procedure. Variations of the fermion variables under space-time varying versions of each of the axial and vector symmetries are generated by

$$\delta_{A,i}^{(n)}(x)\,\psi = I(x)\,\lambda_i\,\gamma_5\,V^n\,\psi$$
$$\delta_{A,i}^{(n)}(x)\,\bar\psi = \bar\psi\,V^{1-n}\,\gamma_5\,\lambda_i\,I(x)$$
$$\delta_{V,i}^{(n)}(x)\,\psi = -i\,I(x)\,\lambda_i\,V^n\,\psi$$
$$\delta_{V,i}^{(n)}(x)\,\bar\psi = i\,\bar\psi\,V^n\,\lambda_i\,I(x)$$
(18)

where $(I(x))_{y,z} = \delta_{y,x}\delta_{x,z}$ denotes the matrix that projects onto site $x$. It is convenient to introduce an infinitesimal function of lattice site, which we call $\varepsilon(x)$ and let $E$ to be the diagonal matrix

$$E = \sum_x \varepsilon(x)\,I(x) \quad (19)$$

We denote the generators of local transformations, weighted by $\varepsilon(x)$, as

$$\Delta_{A,i}^{(n)} = \sum_x \varepsilon(x)\,\delta_{A,i}^{(n)}(x)$$
$$\Delta_{V,i}^{(n)} = \sum_x \varepsilon(x)\,\delta_{V,i}^{(n)}(x)$$
(20)

Their effects on the fermionic action $S_F = \bar\psi(1-V)\psi$ are

$$\Delta_{A,i}^{(n)} S_F = \bar\psi\,\lambda_i\Big[(1-V)\,E\,\gamma_5\,V^n + V^{1-n}\,\gamma_5\,E(1-V)\Big]\psi$$
$$\Delta_{V,i}^{(n)} S_F = -i\,\bar\psi\,\lambda_i\Big[(1-V)\,E\,V^n \;-\; V^n\,E(1-V)\Big]\psi$$
(21)

The conserved currents are the coefficients of $\partial_\mu^{(+)}\varepsilon(x)$ in each of these expressions,

$$\Delta_{A,i}^{(n)} S_F = \sum_{x,\mu}\left(\partial_\mu^{(+)}\varepsilon(x)\right) J_{i\mu}^{5(n)}(x)$$
$$\Delta_{V,i}^{(n)} S_F = \sum_{x,\mu}\left(\partial_\mu^{(+)}\varepsilon(x)\right) J_{i\mu}^{(n)}(x)$$
(22)

where $\partial_\mu^{(+)}$ is the forward, nearest neighbor ordinary (as opposed to covariant) difference.

Note that in Eq. (21) the $n = 0$ and $n = 1$ terms are equal,

$$\Delta^{(0)}_{A,i} S_F = \Delta^{(1)}_{A,i} S_F$$
$$\Delta^{(0)}_{V,i} S_F = \Delta^{(1)}_{V,i} S_F \tag{23}$$

Therefore, even before explicitly constructing the currents, we can conclude that the $n=0$ and $n=1$ currents must be equal.

$$J^{5(0)}_{i\mu}(x) = J^{5(1)}_{i\mu}(x)$$
$$J^{(0)}_{i\mu}(x) = J^{(1)}_{i\mu}(x) \tag{24}$$

In canonically quantized field theory such a mismatch between symmetries and currents could never occur, because in the canonical formulation the operators that generate symmetry transformations are the space integrals of the time components of their conserved currents. However, the possibility for such a situation to arise is inherent in fermionic path integrals, precisely because in Euclidean space path integrals, fermionic and anti-fermionic variables must be treated as independent, not conjugate variables.

To display the currents we will need to use an explicit formulation of lattice chiral fermions, and overlap fermions based on the Wilson fermion action [4] are the most convenient. The overlap kernel is

$$D_{ov} = 1 + \frac{D_W}{\sqrt{D_W^\dagger D_W}} \tag{25}$$

where $D_W$ is the Wilson kernel with a negative mass term

$$D_W = \sum_\mu \frac{1}{2a_\mu} \left[ \gamma_\mu \left(U_\mu - U_\mu^\dagger\right) + \left(2 - U_\mu - U_\mu^\dagger\right) \right] - s \tag{26}$$

Here $(U_\mu)_{x,y} = \delta_{x+\hat{\mu},y} U_\mu(x)$ is a matrix formed from the lattice link variables $U_\mu(x)$ that connect site $x$ to site $x+\hat{\mu}$, $a_\mu$ is the lattice spacing in the $\mu$ direction, and, to eliminate the fermion doublers, the parameter $s$ must satisfy

$$0 < s < \frac{2}{\max_\mu a_\mu} \qquad (27)$$

To evaluate the currents, we first put the site-dependent variations of the action (21) into a form in which the matrix $E$ appears only in commutators.

$$\begin{aligned}
\Delta_{A,i}^{(n)} S_F &= \bar{\psi} \gamma_5 \lambda_i \left\{ [E, V^{n-1}](V-1) - V^{-1}[E,V]V^{n-1} \right\} \psi \\
\Delta_{V,i}^{(n)} S_F &= i \bar{\psi} \lambda_i \left\{ V[E, V^{n-1}](V-1) - [E,V]V^{n-1} \right\} \psi
\end{aligned} \qquad (28)$$

We then expand the commutators using

$$\begin{aligned}
[E,V^{n-1}] &= \sum_{m=0}^{n-2} V^m [E,V] V^{n-m-2} \qquad (n \geq 2) \\
&= -\sum_{m=1}^{1-n} V^{-m} [E,V] V^{n+m-2} \qquad (n \leq 0)
\end{aligned} \qquad (29)$$

which converts the expressions for the variations of the action to sums of terms in which $E$ enters only through the single commutator $[E,V]$. This commutator was evaluated in the original paper of Kikukawa and Yamada [9]. When put into matrix notation, their result reads

$$[E,V] = \sum_{x,\mu} \left( \partial_\mu^{(+)} \varepsilon(x) \right) K_\mu(x) \qquad (30)$$

where

$$K_\mu(x) = \frac{1}{\pi} \int_{-\infty}^{+\infty} \frac{dt}{t^2 + D_W D_W^\dagger} \left[ T_\mu(x) t^2 + D_W T_\mu^\dagger(x) D_W \right] \frac{1}{t^2 + D_W^\dagger D_W}$$

$$T_\mu(x) = \frac{1}{2} \left[ (\gamma_\mu - 1) I(x) U_\mu + (\gamma_\mu + 1) U_\mu^\dagger I(x) \right]$$

(31)

Since the currents are the coefficients of $\partial_\mu^{(+)} \varepsilon(x)$ in $\Delta_{A,V,i}^{(n)} S_F$ (Eq. (22)), they are obtained from the expanded forms of Eq. (28) simply by replacing each occurrence of $[E,V]$ by $K_\mu(x)$. This gives the rather ungainly results

$$J_{i\mu}^{5(n)}(x) = \bar\psi \gamma_5 \lambda_i \left[ \sum_{m=0}^{n-2} V^m K_\mu(x) V^{n-m-2}(V-1) - V^{-1} K_\mu(x) V^{n-1} \right] \psi \quad (n \geq 2)$$

$$= \bar\psi \gamma_5 \lambda_i \left[ -\sum_{m=1}^{1-n} V^{-m} K_\mu(x) V^{n+m-2}(V-1) - V^{-1} K_\mu(x) V^{n-1} \right] \psi \quad (n \leq 0)$$

$$J_{i\mu}^{(n)}(x) = i\bar\psi \lambda_i \left[ V \sum_{m=0}^{n-2} V^m K_\mu(x) V^{n-m-2}(V-1) - K_\mu(x) V^{n-1} \right] \psi \quad (n \geq 2)$$

$$= i\bar\psi \lambda_i \left[ -V \sum_{m=1}^{1-n} V^{-m} K_\mu(x) V^{n+m-2}(V-1) - K_\mu(x) V^{n-1} \right] \psi \quad (n \leq 0)$$

(32)

As noted above, the $n=1$ currents are equal to the $n=0$ currents..

## V. EXTENDED SYMMETRY WARD IDENTITIES

The Euclidean space analogues of conservation laws are the Ward identities. They express the physical implications of symmetries. The Ward identities for each Green's function are derived by making an infinitesimal change of integration variables

$$\psi \to \psi' = \left(1 + \sum_x \varepsilon(x)\, \delta^{(n)}_{AV,i}(x)\right) \psi$$

$$\bar{\psi} \to \bar{\psi}' = \left(1 + \sum_x \varepsilon(x)\, \delta^{(n)}_{AV,i}(x)\right) \bar{\psi} \tag{33}$$

in the path integral expression for that Green's function.

$$\langle \psi \psi \psi \cdots \bar{\psi}\bar{\psi}\bar{\psi} \rangle = \int \mathcal{D}\psi \mathcal{D}\bar{\psi}\; \psi\psi\psi \cdots \bar{\psi}\bar{\psi}\bar{\psi}\, e^{-S_F} \tag{34}$$

[In this discussion we suppress all space-time labels on the fermionic variables and the overall normalization.] Three sorts of terms result from the changes of variable. Besides the explicit action of the $\delta^{(n)}_{AV,i}(x)$ acting on $\psi$ and $\bar{\psi}$, Eq. (18), there are the shifts in the action, Eq. (22),

$$\left(1 + \sum_x \varepsilon(x)\, \delta^{(n)}_{A,i}(x)\right) e^{-S_F} = \left(1 - \sum_{x,\mu} \left(\partial^{(+)}_\mu \varepsilon(x)\right) J^{5(n)}_{i\mu}(x)\right) e^{-S_F}$$

$$\left(1 + \sum_x \varepsilon(x)\, \delta^{(n)}_{V,i}(x)\right) e^{-S_F} = \left(1 - \sum_{x,\mu} \left(\partial^{(+)}_\mu \varepsilon(x)\right) J^{(n)}_{i\mu}(x)\right) e^{-S_F} \tag{35}$$

and also the Jacobians coming from the changes of variable

$$\left.\frac{\partial(\psi',\bar{\psi}')}{\partial(\psi,\bar{\psi})}\right|_{A,i} = \det\left(1 + \sum_x \varepsilon(x)\, I(x)\, \gamma_5\, \lambda_i\, V^n\right) \det\left(1 + \sum_x \varepsilon(x)\, V^{1-n}\, \gamma_5\, \lambda_i\, I(x)\right)$$

$$= 1 + \delta_{i,0} \sum_x \left(\mathrm{Tr}\, \varepsilon(x) I(x) \gamma_5 \left(V^n + V^{1-n}\right)\right) \tag{36}$$

$$\left.\frac{\partial(\psi',\bar{\psi}')}{\partial(\psi,\bar{\psi})}\right|_{V,i} = \det\left(1 - i\sum_x \varepsilon(x)\, I(x)\, V^n\right) \det\left(1 + i\sum_x \varepsilon(x)\, V^n\, I(x)\right) = 1$$

Changes of integration variable do not affect the value of an integral, so the sum of all these first order shifts must vanish. Collecting the coefficient of $\varepsilon(x)$ in the sum gives the Ward identities:

$$\left(\delta_{A,i}^{(n)}(x) - \delta_{i,0} q_A^{(n)}(x)\right)\langle \psi\psi\psi\cdots\bar{\psi}\bar{\psi}\bar{\psi}\rangle + \partial_\mu^{(-)} \langle \psi\psi\psi\cdots\bar{\psi}\bar{\psi}\bar{\psi} J_{i\mu}^{5(n)}(x)\rangle = 0$$
$$\delta_{V,i}^{(n)}(x)\langle \psi\psi\psi\cdots\bar{\psi}\bar{\psi}\bar{\psi}\rangle + \partial_\mu^{(-)} \langle \psi\psi\psi\cdots\bar{\psi}\bar{\psi}\bar{\psi} J_{i\mu}^{(n)}(x)\rangle = 0 \qquad (37)$$

Here we have denoted the contribution coming from the Jacobean of the $\delta_{A,0}^{(n)}(x)$ transformation as

$$q_A^{(n)}(x) = Tr\, I(x)\gamma_5 \left(V^n + V^{1-n}\right) \qquad (38)$$

This is the path integral form of the axial anomaly [15]. All the flavor singlet axial symmetries are anomalous. Note that $q_A^{(n)}(x) = q_A^{(1-n)}(x)$, but that generically the anomalies are different from one another, because they come from the Jacobeans of different transformations $\delta_{A,0}^{(n)}(x)$. By contrast, the integrated anomalies, $\sum_x q_A^{(n)}(x)$, are all equal to the index of the Ginsparg-Wilson kernel, $\sum_x q_A^{(1)}(x)$.

The basic Ward identities come from the fermion propagator $\langle \psi\bar{\psi}\rangle$. They are

$$\gamma_5 \lambda_i I(x) V^n \langle \psi\bar{\psi}\rangle + \langle \psi\bar{\psi}\rangle V^{1-n} I(x) \lambda_i \gamma_5$$
$$+ \partial_\mu^{(-)} \langle \psi J_{i\mu}^{5(n)}(x) \bar{\psi}\rangle + \delta_{i,0} q_A^{(n)}(x) \langle \psi\bar{\psi}\rangle = 0 \qquad (39)$$
$$-i\lambda_i I(x) V^n \langle \psi\bar{\psi}\rangle + i\langle \psi\bar{\psi}\rangle V^n I(x) \lambda_i + \partial_\mu^{(-)} \langle \psi J_{i\mu}^{(n)}(x) \bar{\psi}\rangle = 0$$

For higher Green's functions, the $\delta_{AV}^{(n)}(x)\langle \psi\psi\psi\cdots\bar{\psi}\bar{\psi}\bar{\psi}\rangle$ term expands into a sum of terms, one for each fermion variable, like the first two on the left hand side of (39). The form of these identities, especially those for higher Green's functions, shows the connection with the conservation laws in canonically quantized field theory.

## VI.  AN INVARIANT SUBGROUP AND ITS COSETS

In this section we will study an invariant subgroup of the extended chiral group and the group formed by its cosets. To fix notation, we denote the chiral group by $G$ and the subgroup (to be identified below) by $\tilde{G}$. Generic elements of each are denoted as $g \in G$ and $\tilde{g} \in \tilde{G}$. The condition that the subgroup $\tilde{G}$ is an invariant subgroup is that for any elements $g \in G$ and $\tilde{g} \in \tilde{G}$, $g\,\tilde{g}\,g^{-1}$ is in the subgroup.

$$g\,\tilde{g}\,g^{-1} \in \tilde{G} \tag{40}$$

Each coset of $G$ with respect to $\tilde{G}$ consists of those elements of $G$ which are related by multiplication by some element of $\tilde{G}$. We will redundantly label the cosets by any of the elements they contain, $g\tilde{G}$. The group element $g'$ is an element of $g\tilde{G}$ if there is some element $\tilde{g} \in \tilde{G}$ such that $g'\,\tilde{g} = g$.

There is a natural multiplication law for cosets under which they form a group. It is

$$\left(g_1\tilde{G}\right)\left(g_2\tilde{G}\right) = (g_1 g_2)\tilde{G} \tag{41}$$

This rule is consistent because Eq. (40) assures that the product of any element of $g_1\tilde{G}$ times any element of $g_2\tilde{G}$,

$$g_1\,\tilde{g}_1\,g_2\,\tilde{g}_2 = g_1 g_2 (g_2^{-1}\tilde{g}_1\,g_2)\,\tilde{g}_2 \in (g_1 g_2)\tilde{G} \tag{42}$$

always lies in the coset $(g_1 g_2)\tilde{G}$. The factor group of cosets is denoted as $G/\tilde{G}$.

It is useful to express the above in terms of the generators of the groups $G$ and $\tilde{G}$, and thereby identify the generators of the group of cosets. Let $\Delta$ and $\tilde{\Delta}$ be the Lie algebras of $G$ and $\tilde{G}$, respectively, and let $\delta$ and $\tilde{\delta}$ be generic elements of the respective algebras. Then $(1+\delta)$ and $(1+\tilde{\delta})$ are generic infinitesimal elements of $G$ and $\tilde{G}$, and applying the criterion for the subgroup $\tilde{G}$ to be an invariant subgroup of $G$ to lowest order,

$$(1+\delta)(1+\tilde{\delta})(1-\delta) = (1+[\delta,\tilde{\delta}]) \tag{43}$$

gives the condition for $\tilde{\Delta}$ to generate an invariant Lie subalgebra of $G$, namely that for all elements of $\Delta$ and $\tilde{\Delta}$,

$$[\delta,\tilde{\delta}] \in \tilde{\Delta} \tag{44}$$

With these preliminaries, let us observe that the elements

$$\begin{aligned}
\tilde{\delta}_{A,i}^{(n)} &\equiv \delta_{A,i}^{(n)} - \delta_{A,i}^{(0)} \\
\tilde{\delta}_{V,i}^{(n)} &\equiv \delta_{V,i}^{(n)} - \delta_{V,i}^{(0)}
\end{aligned} \tag{45}$$

are the generators of an invariant subgroup of the extended chiral symmetry group generated by $\delta_{A,i}^{(n)}$ and $\delta_{V,i}^{(n)}$. Their commutators follow from Eq. (11), and are

$$\begin{aligned}
[\tilde{\delta}_{A,i}^{(n)}, \tilde{\delta}_{A,j}^{(m)}] &= i d_{ijk}\left(\tilde{\delta}_{V,k}^{(n-m)} - \tilde{\delta}_{V,k}^{(m-n)}\right) + f_{ijk}\left(\tilde{\delta}_{V,k}^{(n-m)} + \tilde{\delta}_{V,k}^{(m-n)}\right) \\
&\quad - i d_{ijk}\left(\tilde{\delta}_{V,k}^{(-m)} - \tilde{\delta}_{V,k}^{(m)}\right) - f_{ijk}\left(\tilde{\delta}_{V,k}^{(-m)} + \tilde{\delta}_{V,k}^{(m)}\right) \\
&\quad - i d_{ijk}\left(\tilde{\delta}_{V,k}^{(n)} - \tilde{\delta}_{V,k}^{(-n)}\right) - f_{ijk}\left(\tilde{\delta}_{V,k}^{(n)} + \tilde{\delta}_{V,k}^{(-n)}\right) \\
[\tilde{\delta}_{V,i}^{(n)}, \tilde{\delta}_{A,j}^{(m)}] &= i d_{ijk}\left(\tilde{\delta}_{A,i}^{(m-n)} - \tilde{\delta}_{A,i}^{(m+n)}\right) - f_{ijk}\left(\tilde{\delta}_{A,k}^{(m-n)} + \tilde{\delta}_{A,k}^{(m+n)}\right) \\
&\quad - i d_{ijk}\left(\tilde{\delta}_{A,k}^{(-n)} - \tilde{\delta}_{A,k}^{(n)}\right) + 2 f_{ijk}\tilde{\delta}_{A,k}^{(m)} + f_{ijk}\left(\tilde{\delta}_{A,k}^{(-n)} + \tilde{\delta}_{A,k}^{(n)}\right) \\
[\tilde{\delta}_{V,i}^{(n)}, \tilde{\delta}_{V,j}^{(m)}] &= -2 f_{ijk}\tilde{\delta}_{V,k}^{(n+m)} + 2 f_{ijk}\tilde{\delta}_{V,k}^{(m)} + 2 f_{ijk}\tilde{\delta}_{V,k}^{(n)}
\end{aligned} \quad (46)$$

That the commutator algebra of the $\tilde{\delta}_{VA,i}^{(n)}$ closes shows that they indeed generate a proper subgroup. The commutators of the $\tilde{\delta}_{VA,i}^{(n)}$ and $\delta_{VA,i}^{(n)}$ generators are

$$\begin{aligned}
[\delta_{A,i}^{(n)}, \tilde{\delta}_{A,j}^{(m)}] &= i d_{ijk}\left(\tilde{\delta}_{V,k}^{(n-m)} - \tilde{\delta}_{V,k}^{(m-n)}\right) + f_{ijk}\left(\tilde{\delta}_{V,k}^{(n-m)} + \tilde{\delta}_{V,k}^{(m-n)}\right) \\
&\quad - i d_{ijk}\left(\tilde{\delta}_{V,k}^{(n)} - \tilde{\delta}_{V,k}^{(-n)}\right) - f_{ijk}\left(\tilde{\delta}_{V,k}^{(n)} + \tilde{\delta}_{V,k}^{(-n)}\right) \\
[\delta_{V,i}^{(n)}, \tilde{\delta}_{A,j}^{(m)}] &= i d_{ijk}\left(\tilde{\delta}_{A,i}^{(m-n)} - \tilde{\delta}_{A,i}^{(m+n)}\right) - f_{ijk}\left(\tilde{\delta}_{A,k}^{(m-n)} + \tilde{\delta}_{A,k}^{(m+n)}\right) \\
&\quad - i d_{ijk}\left(\tilde{\delta}_{A,k}^{(-n)} - \tilde{\delta}_{A,k}^{(n)}\right) + f_{ijk}\left(\tilde{\delta}_{A,k}^{(-n)} + \tilde{\delta}_{A,k}^{(n)}\right) \\
[\delta_{A,i}^{(n)}, \tilde{\delta}_{V,j}^{(m)}] &= -i d_{ijk}\left(\tilde{\delta}_{A,i}^{(n-m)} - \tilde{\delta}_{A,i}^{(n+m)}\right) - f_{ijk}\left(\tilde{\delta}_{A,k}^{(n-m)} + \tilde{\delta}_{A,k}^{(n+m)}\right) + 2 f_{ijk}\tilde{\delta}_{A,k}^{(n)} \\
[\delta_{V,i}^{(n)}, \tilde{\delta}_{V,j}^{(m)}] &= -2 f_{ijk}\tilde{\delta}_{V,k}^{(n+m)} + 2 f_{ijk}\tilde{\delta}_{V,k}^{(n)}
\end{aligned} \quad (47)$$

That the right hand sides are elements of $\tilde{\Delta}$ shows that the subgroup is an invariant subgroup.

Let us identify the cosets of the extended chiral group. In an infinitesimal neighborhood of the identity, we can uniquely parameterize a general group element as

$$g = 1 + \sum_{i,n}\left[c_{A,i}^{(n)} \delta_{A,i}^{(n)} + c_{V,i}^{(n)} \delta_{V,i}^{(n)}\right] \quad (48)$$

Let us define the sums of the coefficients of the axial and vector generators in this expansion as

$$C_{A,i} = \sum_n c_{A,i}^{(n)}$$
$$C_{V,i} = \sum_n c_{V,i}^{(n)}$$
(49)

If the parameters $C_{A,i}$ and $C_{V,i}$ are all zero, then from Eq. (45) it follows that $g$ is an element of the invariant subgroup $\tilde{G}$. If $g$ and $g'$ are two infinitesimal group elements for which the parameter sums (49) are equal, then $g\,g'^{-1}$ can be written as

$$g\,g'^{-1} = 1 + \sum_n \left(c_{A,i}^{(n)} - c_{A,i}'^{(n)}\right)\delta_{A,i}^{(n)} + \sum_n \left(c_{V,i}^{(n)} - c_{V,i}'^{(n)}\right)\delta_{V,i}^{(n)}$$
(50)

which, by (49), implies that $g\,g'^{-1}$ is an element of the invariant subgroup $\tilde{G}$, and so $g$ and $g'$ are in the same coset. Thus the coefficients $C_{A,i}$ and $C_{V,i}$ provide a complete and non-redundant parameterization of the cosets that contain group elements in the neighborhood of the identity, $\left(g\tilde{G}\right)(C_{A,i}, C_{V,i})$.

We can use this parameterization to determine the local structure of the group of cosets. Let us take any representative infinitesimal group elements from two cosets

$$g = 1 + \delta = 1 + \sum_{i,n}\left[c_{A,i}^{(n)}\,\delta_{A,i}^{(n)} + c_{V,i}^{(n)}\,\delta_{V,i}^{(n)}\right] \in \left(g\tilde{G}\right)(C_{A,i}, C_{V,i})$$
$$g' = 1 + \delta' = 1 + \sum_{j,m}\left[c_{A,j}'^{(m)}\,\delta_{A,j}^{(m)} + c_{V,j}'^{(m)}\,\delta_{V,j}^{(m)}\right] \in \left(g\tilde{G}\right)(C_{A,j}', C_{V,j}')$$
(51)

and calculate their group commutator $g\,g'\,g^{-1}\,g'^{-1}$. The leading term is $1+[\delta,\,\delta']$, which is evaluated using the commutation relations given in Eq. (11). The result is another infinitesimal group element

$$g'' = 1+[\delta,\,\delta'] = g\,g'\,g^{-1}\,g'^{-1} = 1 + \sum_{k,p} c_{A,k}''^{(p)}\,\delta_{A,k}^{(p)} + c_{V,k}''^{(p)}\,\delta_{V,k}^{(p)}$$
(52)

where the coefficients are

$$c_{A,k}^{\prime\prime(p)} = \sum_{i,j,n,m} \left( i d_{ijk}(\delta_{p,m-n} - \delta_{p,m+n}) - f_{ijk}(\delta_{p,m+n} + \delta_{p,m-n}) \right) c_{V,i}^{(n)} c_{A,j}^{\prime(m)}$$

$$+ \sum_{i,j,n,m} \left( -i d_{ijk}(\delta_{p,n-m} - \delta_{p,m+n}) + f_{ijk}(\delta_{p,m+n} + \delta_{p,n-m}) \right) c_{A,i}^{(n)} c_{V,j}^{\prime(m)}$$

$$c_{V,k}^{\prime\prime(p)} = \sum_{i,j,n,m} \left( i d_{ijk}(\delta_{p,n-m} - \delta_{p,m-n}) + f_{ijk}(\delta_{p,n-m} + \delta_{p,m-n}) \right) c_{A,i}^{(n)} c_{A,j}^{\prime(m)}$$

$$+ \sum_{i,j,n,m} \left( -2 f_{ijk} \delta_{p,n+m} \right) c_{V,i}^{(n)} c_{V,j}^{\prime(m)}$$

(53)

Summing over $n, m,$ and $p$, reduces this to

$$C_{A,k}^{\prime\prime} = 2 \sum_{i,j} f_{ijk} \left( C_{A,i} C_{V,j}^{\prime} - C_{V,i} C_{A,j}^{\prime} \right)$$

$$C_{V,k}^{\prime\prime} = 2 \sum_{i,j} f_{ijk} \left( C_{A,i} C_{A,j}^{\prime} - C_{V,i} C_{V,j}^{\prime} \right)$$

(54)

confirming that the coset in which the group commutator lies, $(g\tilde{G})(C_{A,i}^{\prime\prime}, C_{V,i}^{\prime\prime})$, does not depend on the choice of representative group elements.

From Eq. (51) one sees that the coset parameters label the cosets linearly near the origin, and so we may express a generic coset near the identity element of the group of cosets as

$$(g\tilde{G})(C_{A,i}, C_{V,i}) = 1 + \sum_i \left[ C_{A,i} (\delta\tilde{G})_{A,i} + C_{V,i} (\delta\tilde{G})_{V,i} \right]$$

(55)

and take the above as an implicit definition of the coset group generators $(\delta\tilde{G})_{A,i}$ and $(\delta\tilde{G})_{V,i}$. From (54) and (55) we can read off the commutation relations of the generators

$$\left[ (\delta\tilde{G})_{A,i}, (\delta\tilde{G})_{A,j} \right] = +2 f_{ijk} (\delta\tilde{G})_{V,k}$$

$$\left[ (\delta\tilde{G})_{V,i}, (\delta\tilde{G})_{A,j} \right] = -2 f_{ijk} (\delta\tilde{G})_{A,k}$$

$$\left[ (\delta\tilde{G})_{V,i}, (\delta\tilde{G})_{V,j} \right] = -2 f_{ijk} (\delta\tilde{G})_{V,k}$$

(56)

The behavior of both the invariant subgroup and its factor group under CP follow simply from the CP transformation properties of the generators, Eq. (17). Applying these to the generators of the invariant subgroup (45), we find (in a flavor basis where the matrices $\lambda_i$ are hermitian)

$$(\mathcal{CP})^{-1} \tilde{\delta}_{A,i}^{(n)} (\mathcal{CP}) = -\tilde{\delta}_{A,i}^{(1-n)} + \tilde{\delta}_{A,i}^{(1)}$$
$$(\mathcal{CP})^{-1} \tilde{\delta}_{V,i}^{(n)} (\mathcal{CP}) = -\tilde{\delta}_{V,i}^{(n)} \tag{57}$$

Thus CP is an outer automorphism of the invariant subgroup of the extended chiral group as well as of the full group. As in the case of the full group, CP mixes the axial generators.

The CP transformation properties of the factor group of cosets follow from the action of CP on a general infinitesimal element of the extended chiral group, Eq. (48).

$$(\mathcal{CP})^{-1} g (\mathcal{CP}) = 1 - \sum_{i,n} \left[ c_{A,i}^{(1-n)} \delta_{A,i}^{(n)} + c_{V,i}^{(n)} \delta_{V,i}^{(n)} \right] \tag{58}$$

and so all the elements of each coset are transformed into a single image coset,

$$(\mathcal{CP})^{-1} \left( g\tilde{G} \right)(C_{A,i}, C_{V,i})(\mathcal{CP}) = \left( g\tilde{G} \right)(-C_{A,i}, -C_{V,i}) \tag{59}$$

We can express this result in terms of the symbolic generators of the factor group, as defined in Eq. (55) as

$$(\mathcal{CP})^{-1} \left( \delta\tilde{G} \right)_{A,i} (\mathcal{CP}) = - \left( \delta\tilde{G} \right)_{A,i}$$
$$(\mathcal{CP})^{-1} \left( \delta\tilde{G} \right)_{V,i} (\mathcal{CP}) = - \left( \delta\tilde{G} \right)_{V,i} \tag{60}$$

This result, together with the commutators of Eq. (56), show that the structure of the coset factor group $\left( g\tilde{G} \right)$ is locally identical to continuum chiral symmetry.

The physical meaning of the invariant subgroup can be seen from the Ward identities for the tilde transformations. Each term in $\tilde{\delta}_{AV}^{(n)}(x)\langle\psi\psi\psi\cdots\bar{\psi}\bar{\psi}\bar{\psi}\rangle$ includes a factor of $I(x)(V^n-1)(1-V)^{-1}$ or $I(x)(V^{1-n}-1)(1-V)^{-1}$. Both give $I(x)$ times finite sums of powers of either $V$ or $V^\dagger$; the usual propagator $(1-V)^{-1}$ attached to $I(x)$ has been cancelled. This means that the non-anomalous Green's functions $\langle\psi\psi\psi\cdots\bar{\psi}\bar{\psi}\bar{\psi}\,\tilde{J}_{i\mu}^{5(n)}(x)\rangle$ and $\langle\psi\psi\psi\cdots\bar{\psi}\bar{\psi}\bar{\psi}\,\tilde{J}_{i\mu}^{(n)}(x)\rangle$ will each be given by a sum of fermionic Green's functions, each of which has one external fermion line missing its propagator. Truncating the external lines, *i.e.* multiplying by inverse propagators and going to the mass shell, will therefore give zero for each term. The conclusion of these considerations is that in the continuum limit, no physical states will couple to the vector or to the flavored (non-anomalous) axial currents associated with the $\tilde{\delta}_{AV}^{(n)}$ transformations, that is, with the symmetries in the invariant subgroup.

## VII. CONTINUATION TO MINKOWSKI SPACE

As was noted earlier, the mismatch between the Euclidean space currents and the Euclidean space symmetries — two different symmetry transformations having the same Noether current — could not occur in a Minkowski space canonically quantized field theory, because in the canonical formulation the generators of symmetry transformations are the space integrals of the time components of their conserved currents. This observation indicates that a straightforward analytic continuation of Ginsparg-Wilson fermions to Minkowski must encounter problems. To see how these problems manifest themselves, we will examine the simplest example of Ginsparg-Wilson fermions, overlap fermions without gauge fields. Being able to continue to Minkowski space only requires that the time dimension be continuous, and it will clarify the discussion to allow for the spacings along the space and Euclidean time directions to be different.

The Wilson kernel without gauge fields is then

$$D_W = \sum_\mu \frac{1}{2a_\mu} \left[ \gamma_\mu \left( \partial_\mu^{(+)} + \partial_\mu^{(-)} \right) - \left( \partial_\mu^{(+)} - \partial_\mu^{(-)} \right) \right] - s \tag{61}$$

where $a_4$ is the lattice spacing in the Euclidean time direction and $a_1 = a_2 = a_3 \equiv a_s$ is the lattice spacing in the spatial directions. The negative mass term $s$ is constrained as in (27) to lie between $0$ and $2$ over the largest lattice spacing $a_\mu$. In momentum space, where $D_W$ is diagonal, this gives for the Ginsparg-Wilson kernel

$$a D_{ov} = 1 - V = 1 + \frac{D_W}{\sqrt{D_W^\dagger D_W}} = \frac{i\gamma \cdot \hat{p} + M + \sqrt{\hat{p}^2 + M^2}}{\sqrt{\hat{p}^2 + M^2}} \tag{62}$$

where we have denoted

$$\begin{aligned}
\hat{p}_\mu &= \frac{1}{a_\mu} \sin(a_\mu p_\mu) \\
\hat{\hat{p}}_\mu &= \frac{2}{a_\mu} \sin\left( \frac{a_\mu p_\mu}{2} \right) \\
M &= \left( \sum_{\mu=1}^{4} \frac{a_\mu}{2} \hat{\hat{p}}_\mu^2 \right) - s
\end{aligned} \tag{63}$$

The parameter $a$ with no subscript has dimensions of length and is fixed by the normalization of the fermion propagator. Eq. (62) of course satisfies the Ginsparg-Wilson equation and the superabundance of symmetries and the mismatch between symmetries and currents are both still present. The symmetry generators and Noether currents reduce to Eqs. (9), (32), and (31) with

$$V = -\frac{i\gamma \cdot \hat{p} + M}{\sqrt{\hat{p}^2 + M^2}} \tag{64}$$

Let us keep the spatial lattice intact but take $a_4$ to zero. In this limit the Ginsparg-Wilson kernel becomes

$$a D_{ov} \to a D^{(3)} = \frac{i\gamma_4 \, p_4 + i\gamma^{(3)} \cdot \hat{p}^{(3)} + M^{(3)} + \sqrt{p_4^2 + \hat{p}^{(3)2} + M^{(3)2}}}{\sqrt{p_4^2 + \hat{p}^{(3)2} + M^{(3)2}}} \tag{65}$$

where the superscript $^{(3)}$ on $\gamma$ and $\hat{p}$ connotes the 1, 2, and 3 components and

$$M^{(3)} = \lim_{a_4 \to 0} M = \left(\frac{a_s}{2} \sum_{i=1}^{3} \hat{p}_i^2\right) - s \tag{66}$$

In the $a_4 \to 0$ limit the trig functions of the Euclidean energy are eliminated, $\hat{p}_4 \to p_4$, and the terms involving $\hat{p}_4$ vanish. In this limit the domain of complex energy becomes the full complex plane and the natural expectation would be that one should obtain, after analytic continuation in energy, a quantum mechanical theory of fermions on a spatial lattice. This expectation is not fulfilled.

In fact, the analytic structure of the kernel is inappropriate to a sensible real time theory, because there are unphysical square root branch points at Euclidean energies

$$p_4 = \pm i \sqrt{\hat{p}^{(3)2} + M^{(3)2}} \tag{67}$$

These branch points occur at real (Minkowski) energy. Their presence means that the resulting Dirac kernel is non-local in time. This non-locality is mathematically related to the familiar failure of lattice ultralocality [16], but its meaning is quite different. This is a real, physical non-locality in time, because it has a physically finite temporal extent even in the $a_4 \to 0$ limit. The extent of the non-locality in the continuous time direction is given by Eq. (67), and so it varies with the spatial momentum. The scale the non-locality in time is set by the lattice spacing in the space dimensions, $a_s$. The resulting theory has all the spurious extra symmetries of discrete time Ginsparg-Wilson fermions, and so some impediment to reaching a sensible real time lattice theory had to appear. The unphysical singularities at the locations given by Eq, (67) are the expected impediments.

In contrast with other Euclidean lattice actions, where taking the continuous-time limit directly gives the canonical Hamiltonian [17, 18], there is no Hamiltonian corresponding to the $a_4 \to 0$ limit of free overlap fermions.

The fact that taking the continuous time limit of overlap fermions does not result in a Hamiltonian theory, and the fact that the overlap kernel is at best exponentially local in the continuum, are both manifestations of the fact that overlap fermions do not satisfy reflection positivity. If they did, that would assure that straightforward analytic continuation of the Euclidean space Green's functions to Minkowski space would give the target Minkowski space theory [19].

The failure of the continuum time limit alone to give a sensible theory does not mean that fully space-time continuous chiral fermions are out of reach. If we also take the limit of continuous space, $a_s \to 0$, then $\hat{p}^{(3)} \to p^{(3)}$, $M^{(3)} \to -s$, and the free Ginsparg-Wilson kernel becomes

$$a D_{ov} \to a D^{(cont)} = \frac{i\gamma_4 \, p_4 + i\gamma^{(3)} \cdot p^{(3)} - s + \sqrt{p_4^2 + p^{(3)2} + s^2}}{\sqrt{p_4^2 + p^{(3)2} + s^2}} \tag{68}$$

This is still a non-local theory, because of the singularities at

$$p_4 = \pm i\sqrt{p^{(3)2} + s^2} \tag{69}$$

However, if the negative mass parameter $s$ is taken proportional to the inverse lattice spacing and the action is normalized by taking $a s = 1$, these spurious singularities are banished to infinity and we recover the free massless continuum Dirac kernel

$$D_{ov} \to i\gamma \cdot p \tag{70}$$

There is no impediment to analytically continuing this theory to Minkowski apace. This Dirac kernel, of course, admits no extra symmetries. In place of the Ginsparg-Wilson relation it satisfies ordinary chiral symmetry. At the end of this sequence of limits, as can be seen from Eq. (64),

$$V \to \lim_{\substack{a_\mu \to 0 \\ s \to \infty}} \left( -\frac{i\gamma \cdot \hat{p} + M}{\sqrt{\hat{p}^2 + M^2}} \right) = 1 \tag{71}$$

The sequence of limits that were required to produce a valid quantum theory from free overlap fermions also eliminated all remnants of lattice Ginsparg-Wilson chirality.

## VIII. CONJUGATE FERMION TRANSFORMATIONS

In this article, we have discussed several peculiar properties of the chiral symmetry of Ginsparg-Wilson lattice chiral fermions. The specific issues we uncovered were:

1) The chiral symmetry group of Ginsparg-Wilson lattice fermions is not the same as the continuum chiral symmetry group. There are an infinite number of lattice chiral transformations corresponding to each continuum transformation, and in consequence the lattice theory has an unphysical superabundance of conserved quantities

2) The connection between conserved charges and symmetry generators is (partially) lost. The generators of two pairs of different transformations, ($\delta_{A,i}^{(1)} \neq \delta_{A,i}^{(0)}$ and $\delta_{V,i}^{(1)} \neq \delta_{V,i}^{(0)}$) have the same Noether currents ($J_{i\mu}^{5(0)} = J_{i\mu}^{5(1)}$ and $J_{i\mu}^{(0)} = J_{i\mu}^{(1)}$). This is obviously incompatible with canonically quantized field theory, where the generator of a symmetry transformation is the space integral of the time component of its conserved current.

This situation could arise because in the Euclidean path integral the anti-fermions are described by variables that are not conjugate to the fermion variables. But why does this phenomenon, clearly a latent possibility in every Euclidean path integral treatment of a field theory, actually arise only in consideration of Ginsparg-Wilson fermions?

The difference between Ginsparg-Wilson chirality and other symmetries analyzed by Euclidean path integrals is that while fermionic and anti-fermionic variables always enter

Euclidean space path integrals as independent, not conjugate variables, we ordinarily consider only symmetry transformations on $\psi$ and $\bar{\psi}$ that are compatible with their being each others conjugates. However, the axial symmetry of the Ginsparg-Wilson action discovered by Lüscher, and its generalizations discussed here, are incompatible with $\psi$ and $\bar{\psi}$ being conjugate variables.

Specifically, under the transformations generated by $\delta_{A,i}^{(n)}$, the Euclidean Dirac conjugates of the $\bar{\psi}$ and $\psi$ variables transform as

$$\begin{aligned} \delta_{A,i}^{(n)}\left(\gamma_5 \bar{\psi}^\dagger\right) &= \lambda_i \gamma_5 V^{1-n}\left(\gamma_5 \bar{\psi}^\dagger\right) \\ \delta_{A,i}^{(n)}\left(\psi^\dagger \gamma_5\right) &= \left(\psi^\dagger \gamma_5\right) V^n \gamma_5 \lambda_i \end{aligned} \qquad (72)$$

These are not the same as the $\delta_{A,i}^{(n)}$ transformations on $\psi$ and $\bar{\psi}$. In fact, they are the transformations generated by $\delta_{A,i}^{(1-n)}$.

This observation suggests a simple solution to all the problems of lattice chiral symmetry we have been analyzing: Generalize the $\delta_{A,i}^{(n)}$ transformations non-integral $n$, and take the group generated by $\delta_{A,i}^{(1/2)}$ and $\delta_{V,i}^{(0)}$ as the lattice chiral symmetry group. According to Eq. (11), the algebra of this group is closed and requires no additional symmetries. It is exactly the same as the continuum chiral symmetry group. Furthermore, according to Eq. (17), these generators are eigenvectors of CP with the correct eigenvalues.

Alas, this "solution" is a mirage.

The reason is that the matrix $V^{1/2}$ must be singular. A simple way to see this is to note that while the eigenvalues of $V$ lie anywhere on the unit circle in the complex plane, the eigenvalues of $V^{1/2}$ are restricted to the right half of the unit circle. Something must go wrong when an eigenvalue of $V$ approaches $-1$. The free overlap fermion example discussed in the previous section illustrates this in detail.

For this example, $V$ is given by Eq. (64), and its square root is

$$V^{1/2} = \frac{-i\sqrt{\frac{M+\sqrt{\hat{p}^2+M^2}}{2\hat{p}^2}}\,\gamma\cdot\hat{p} + \sqrt{\frac{-M+\sqrt{\hat{p}^2+M^2}}{2}}}{\sqrt[4]{\hat{p}^2+M^2}} \qquad (73)$$

At all 16 corners of the Brillouin zone, $V$ is well defined. At the origin, $p=(0,0,0,0)$, $M$ is negative and $V$ is the unit matrix. At the other 15 corners, $M$ is positive and $V$ is the negative of the unit matrix. At the origin, $V^{1/2}$ is also perfectly well defined and is also the unit matrix. However, in the neighborhoods of the other 15 corners of the Brillouin zone, $V^{1/2}$ is ill defined. Specifically, as $p$ approaches any corner of the Brillouin zone except the origin,

$$V^{1/2} \rightarrow -i\frac{\gamma\cdot\hat{p}}{\sqrt{\hat{p}^2}} \qquad (74)$$

The result is finite no matter how one takes the limit, and its eigenvalues, $\pm 1$, are also fixed. However, the matrix itself, and so which eigenvectors have which eigenvalues, depends entirely on the direction from which the corner is approached.

## IX. CONCLUSIONS AND SUMMARY

In this article, we have analyzed the group of vector and axial vector symmetries of Ginsparg-Wilson lattice chiral fermions. We found that the symmetry group on the lattice is not the same as the continuum chiral symmetry group. There are an infinite number of lattice chiral transformations corresponding to each continuum transformation. We also showed that the lattice chiral group has an infinite parameter invariant subgroup which encapsulates the "extra" symmetry transformations. Its factor group is isomorphic to the continuum chiral symmetry group. However, the factor group is not represented on the fermionic variables, and so it cannot be used in any straightforward way to represent continuum chiral symmetry on the lattice.

We also constructed the Noether currents corresponding to the generators of the extended lattice group, and derived their Ward identities. In so doing we found a peculiar situation, namely that the connection between conserved charges and symmetry generators was (partially) lost. There are two pairs of different transformations whose generators, $\delta_{A,i}^{(1)} \neq \delta_{A,i}^{(0)}$ and $\delta_{V,i}^{(1)} \neq \delta_{V,i}^{(0)}$, have the same Noether currents, $J_{i\mu}^{5(0)} = J_{i\mu}^{5(1)}$ and $J_{i\mu}^{(0)} = J_{i\mu}^{(1)}$. We noted that this situation is incompatible with canonically quantized field theory, where the generator of a symmetry transformation is the space integral of the time component of its conserved current.

This situation could arise because the Euclidean path integral does not automatically give a canonical field theory, since anti-fermions are described by variables that are not the conjugates of the fermion variables. The reason this situation actually arose is not just a result of the independence of fermion anti-fermion variables, but because all of the axial symmetry transformations generated by the $\delta_{A,i}^{(n)}$ are incompatible with the anti-fermion variables being the conjugates of the fermion variables.

We examined the lattice chiral symmetry group in the solvable example of free overlap fermions, paying attention to the limit in which Euclidean time is taken continuous but the spatial lattice remains discrete. The continuous time limit retains all the extra symmetry of the fully latticized theory, and there are also different symmetries that have the same Noether generator. We observed that there were singularities in the complex energy plan that prevented the construction of a Hamiltonian theory, as could be expected since the continuous time limit by itself leaves intact the features of Ginsparg-Wilson fermions that are in contradiction with canonical quantization.

These properties of Ginsparg-Wilson lattice fermions form a coherent picture. The singularities are the mechanism by which symmetries that are incompatible with canonical field theory, such as the axial symmetries of Ginsparg-Wilson fermions, are blocked from being analytically continued from Euclidean to Minkowski space.

All of these considerations were uncovered by analyzing the symmetry group of lattice fermions satisfying the Ginsparg-Wilson equation. As such, these features are quite general and are inherent in all implementations of lattice fermions based on the Ginsparg-Wilson relation, including overlap, domain-wall, and perfect-action chiral fermions.

**Acknowledgments**

The author would like to thank Weonjong Lee, Steve Sharpe, and Larry Yaffe for discussions and advice about this work, and Wolfgang Bietenholz, Daniel Nogradi, and an anonymous reviewer for comments on ref [7].